%% file: AUMAC_arXiv.tex
\documentclass[10pt, conference, letter]{IEEEtran}%% Depending on your installation, you may wish to adjust the top margin:
 \usepackage{microtype}
\usepackage[utf8]{inputenc} 
\usepackage[T1]{fontenc}
\usepackage{url}              % provides \url{...}
\usepackage{ifthen}           % provides \ifthenelse
\usepackage{cite}             % improves presentation of citations
\usepackage[cmex10]{amsmath}  % Use the [cmex10] option to ensure compliance
\usepackage{adjustbox}
\usepackage{amssymb,amsfonts}
\usepackage{bbm}
\usepackage{dsfont}
\usepackage{mleftright}       % fix to wrong spacing of \left-,
\mleftright                   % \middle- \right-commands 
\usepackage{graphicx}         % provides \includegraphics{...} to
\usepackage{booktabs}         % fixes poor spacing in tables and
\usepackage{pgf}
\usepackage{caption}
\usepackage{lipsum} %for writing one-column equations in a two-column document with \begin{widetext}
\usepackage{cuted}
\usepackage{stfloats}
\usepackage{etoolbox}
\usepackage[switch]{lineno} 
\usepackage{pgfplots}
\pgfplotsset{width=0.8\linewidth,compat=1.9}
\usepackage{setspace}
\newcommand{\pow}{\mathsf{P}}

\newcommand{\argmin}[1]{\ensuremath{\underset{#1}{\arg\min}}}
\newcommand{\arginf}[1]{\ensuremath{\underset{#1}{\arg\inf}}}

\newcommand{\sett}{\mathcal S}

\newcommand{\setd}{d_{[\ka]}\!}

\newcommand{\ka}{\ensuremath{\normalfont \textsf{K}_{\text{a}}}}

\newcommand{\dm}{\ensuremath{\textsf{D}_{\text{m}}}}

\newcommand{\textsfM}{{\normalfont \textsf{M}}}

\newcommand{\logM}{{\normalfont \log\textsf{M}}}

\newcommand{\ind}{\ensuremath{\mathbbm{1}}}

\newcommand{\ebno}{\ensuremath{\frac{\text{E}_{\text{b}}}{\text{N}_{\text{0}}}}}

\newcommand{\matr}[1]{\mathbf{#1}}

\renewcommand{\baselinestretch}{1}
\newtheorem{thm}{Theorem}
\newtheorem{lem}{Lemma}
\newtheorem{cor}{Corollary}

\newtheorem{defi}{Definition}

\newtheorem{rmk}{Remark}

\newenvironment{theorem}{\begin{thm}}{\vspace{0cm}\end{thm}}
\newenvironment{definition}{\begin{defi}}{\end{defi}}
\newenvironment{lemma}{\begin{lem}}{\end{lem}}

\begin{document}
\title{Wrap-Decoding in Asynchronous Unsourced Multiple Access With and Without Delay Information} 

%%%%%%
\author{%
  \IEEEauthorblockN{Jyun-Sian Wu, Pin-Hsun Lin, Marcel A. Mross, and Eduard A. Jorswieck}\\
  \IEEEauthorblockA{Institute for Communications Technology, Technische Universität Braunschweig, Germany\\
  \{jyun-sian.wu, p.lin, m.mross, e.jorswieck\}@tu-braunschweig.de
  % p.lin@tu-braunschweig.de
    }
  %   \and
  %   \IEEEauthorblockN{Pin-Hsun Lin}
  % \IEEEauthorblockA{Technical University of Braunschweig\\
  % p.lin@tu-braunschweig.de
  %   }
  %   \and
  %   \IEEEauthorblockN{Marcel Mross}
  % \IEEEauthorblockA{Technical University of Braunschweig\\
  % m.mross@tu-braunschweig.deC.G.F. }
  %   \and
  %   \IEEEauthorblockN{Eduard Jorswieck}
  % \IEEEauthorblockA{Technical University of Braunschweig\\
  % e.jorswieck@tu-braunschweig.de
  %   }
}

\maketitle

%%%%%
%% Abstract: 
%% If your paper is eligible for the student paper award, please add
%% the comment "THIS PAPER IS ELIGIBLE FOR THE STUDENT PAPER
%% AWARD." as a first line in the abstract. 
%% For the final version of the accepted paper, please do not forget
%% to remove this comment!
%%
\begin{abstract}
     An asynchronous $\ka$-active-user unsourced multiple access channel (AUMAC) is a key model for uncoordinated massive access in future networks. We focus on a scenario where each transmission is subject to the maximal delay constraint ($\dm$), yet the precise delay of each user is unknown at the receiver. The combined effects of asynchronicity and uncertain delays require analysis over all possible delay-codeword combinations, making the complexity of the analysis grow with $\dm$ and $\ka$ exponentially.
    To overcome the complexity, we employ a wrap-decoder for the AUMAC and derive a uniform upper bound on the per-user probability of error (PUPE). Numerical results illuminate the trade-off between energy per bit and the number of active users under various delay constraints. Furthermore, in our considered AUMAC, decoding without explicit delay information is shown to achieve nearly the same energy efficiency as decoding with perfect delay knowledge.
\end{abstract}

\input{Contents/Introduction}

\input{Contents/SystemModel}

\input{Contents/Main_result_arXiv}
\input{Contents/Numerical}

\input{Contents/Conclusion}
\appendices
\input{back/Appendix_Proof_thm1}

\input{back/Appendix_Proof_thm2}

\input{back/Appendix_Proof_lemma}
\bibliographystyle{IEEEtran}
% \enlargethispage{-12cm}
\clearpage
\IEEEtriggeratref{10}
% \IEEEtriggeratref{4} \newpage 
 \newpage 
\bibliography{references}

%%
%% where we here have assumed the existence of the files
%% definitions. Bib and bibliofile. Bib.
%% BibTeX documentation can be obtained at:
%% http://www.ctan.org/tex-archive/biblio/bibtex/contrib/doc/
%%%%%%
%% Or you use manual references (pay attention to consistency and the
%% formatting style!):

%%%%%% 
%% Appendix:
%% If needed, a single appendix is created by
%%
% \newpage

% \input{back/Appendix_C}

%%
%% If several appendices are needed, then the command
%%
% \appendices
%%
%% in combination with further \section-commands can be used.
%%%%%%

\end{document}

%% file: Contents/Introduction.tex
% !TeX root = ../ISIT2024.tex

\section{Introduction}

Massive access techniques have attracted significant attention in 6G and beyond, especially for Internet-of-things, sensor networks, and massive machine-type communication \cite{lin2023information}. A key challenge lies in designing short-blocklength codebooks to enable numerous devices to access an access point simultaneously. Conventional multiple-access channel (MAC) systems, which rely on individual codebooks per device \cite{ElGamal.2011}, are impractical for such scenarios. To address this, the \textit{unsourced multiple-access channel} (UMAC) was proposed \cite{Polyanskiy_perspective_on_UMAC}, where all transmitters share an identical codebook.

Key results on the UMAC include asymptotic capacity analysis for user numbers scaling with blocklength \cite{Guo_many_access_asymptotic_capacity}, second-order asymptotic achievable rates in grant-free random access \cite{Effros_RAC, Effros_RAC_MAC}, and energy efficiency under per-user probability of error (PUPE) constraints \cite{Polyanskiy_perspective_on_UMAC}. A practical T-fold ALOHA scheme for grant-free Gaussian random access has also been proposed, with energy efficiency analysis \cite{T_fold_1}.

Asynchronous systems are worth investigating due to the difficulty of synchronizing a large number of devices. For asynchronous MAC, asymptotic capacity matches synchronous MAC if the delay-to-blocklength ratio asymptotically vanishes \cite{AMUC1981,Cyclic_decoding}. For asynchronous UMAC (AUMAC), authors in \cite{Polyanskiy_low_complexity_AUMAC, Polyanskiy_Short_packet} consider T-fold ALOHA \cite{T_fold_1} in orthogonal frequency-division multiplexing (OFDM), where the length of the cyclic prefix must be greater than the maximum delay. Sparse orthogonal frequency-division multiple access (OFDMA) and compressed sensing techniques are applied for identifying devices and decoding codewords \cite{Guo_CS}. A joint detection of delay and user activity is formulated by tropical linear algebra~in~\cite{Group_testing_delays}.

The PUPE can be bounded by considering all possible incorrectly decoded codewords transmitted by the users in subsets $\sett \subseteq [\ka]$. For UMAC, the permutation-invariant property \cite{Polyanskiy_perspective_on_UMAC} guarantees that for a given number of incorrectly decoded codewords, $|\sett|$, the probabilities that the receiver incorrectly decodes all combinations of $|\sett|$ out of $\ka$ transmitted codewords are identical. This permutation-invariance, however, is destroyed by the AUMAC, where different combinations of codewords and delays are distinguishable. Therefore, even though the receiver is assumed to have perfect delay information, the PUPE of AUMAC consists of many more different error events that must be considered, leading to a more complex~error~bound.

To address the challenging analysis of AUMAC, we investigated the worst-case delay under the assumption that the receiver decodes at the $n$-th channel use with perfect delay information in \cite{ISIT_worst_case}. In contrast, this work introduces a fundamentally different scenario by considering the AUMAC with the receiver decoding the transmitted codewords after receiving $n+\dm$ symbols and decoding without delay information, where $\dm$ represents the maximal delay constraint. While in \cite{ISIT_worst_case}, we were able to find a uniform upper bound by finding the worst-case delay, finding a uniform upper bound with respect to the worst-case delay in our new model is much more challenging.

Therefore, inspired by \cite{Cyclic_decoding}, we apply the wrap-decoding strategy to AUMAC: the receiver superimposes the last $\dm$ received symbols and the first $\dm$ received symbols. As a result, the decoder observes the sum of $\ka$ cyclically shifted codewords plus Gaussian noise, where the noise power in the first $\dm$ symbols is twice that of the remaining symbols. For AWGN channels and i.i.d. Gaussian codewords, the wrap-decoder ensures that the PUPE remains identical for all delays, such that finding the worst case becomes unnecessary. We can also analyze the PUPE of the AUMAC for the cases with and without perfect delay information at the receiver since obtaining the perfect delay information is difficult for the receiver in practical massive access systems. Although evaluating the PUPE of the AUMAC without delay information is more difficult compared to the AUMAC with delay information, the wrap-decoding enables us to derive a uniform upper bound for all codeword-delay combinations. Consequently, the PUPE of both cases can be derived by scaling the uniform upper bounds with the number of the corresponding error~events.

Furthermore, the numerical result illustrates the improvement over \cite{ISIT_worst_case} due to decoding after $n+\dm$ received symbols. We also show that the wrap-decoder design not only enhances robustness against different delays but also against the lack of delay information, as the performance of the wrap-decoder remains nearly identical with and without delay information.

% =======================================================
\textit{Notation:} We will denote $f^{(i)}(t)$ as the $i$-th derivative of $f(x)$ at the point $x=t$ and $f_{1,y}^{(i)}(x,t)$ as the $i$-th partial derivative of $f_1(x,y)$ with respect to  $y$ at the point $y=t$. We use the indicator function $\ind(\cdot)$, the natural logarithm $\log(\cdot)$, and the Landau symbol $O(\cdot)$. The binomial coefficient of $n$ out of $k$ is represented by $\binom{n}{k}$. The number of permutations of $k$ is denoted as $k!$. The transpose of a real matrix $\matr{A}$ is denoted as $\matr{A}^\top.$ We use $\|\matr{A}\|_1$ to denote the 1-norm of the matrix $\matr{A}$.We denote $[k]=\{1,2,...,k\}$, $X_{[k]}:=\{X_1,X_2,...,X_k\}$, and $\mathcal F\!\setminus\!\mathcal T\!=\!\{x:x\in \mathcal F, x \!\not\in\!\mathcal T\}$, where $\mathcal F$ and $\mathcal T$ are two sets. 
% For any set $\mathcal F=\{F_1,F_2,...,F_{|\mathcal F|}\}$, we denote $\{X_m\}_{m\in \mathcal F}=\{X_{F_1},X_{F_2},...,X_{F_{|\mathcal F|}}\}$.
% \vspace{-1mm}

% %Here ends the 2D plot

%% file: Contents/SystemModel.tex
% !TeX root = ../ISIT2024.tex
% \vspace{-0.2cm}
\section{System model and preliminaries} \label{sec: system model} 
We consider an AUMAC with $\ka$ active transmitters, one receiver, and additive white Gaussian noise (AWGN). All transmitters use the same codebook with the same maximal power constraint, $\pow'$, to transmit a fixed $\log_2\textsfM$ bits payload. The codewords are independent and identically distributed (i.i.d.) generated from a Gaussian distribution with mean zero and variance $\pow$, where $\pow<\pow'$. The power backoff reduces the probability that the maximal power constraint is violated.

\begin{definition}\label{def: delay set}
    We define the asynchronicity with delay constraint $\dm$ by a vector of delays as $$    \setd:=[d_1,\ d_2,\ ..., \ d_{\ka }] \in \{\mathbb N_0\}^{\ka},$$
    where $d_i$ represents the delay of the $i$-th received codeword, $d_{i}\leq d_{i+1}, \; \forall i\in[\ka-1]$ and $d_{\ka}\leq \dm$. We define $\alpha:= \frac{\dm}{n}\in[0,1)$, which is a constant with respect to the blocklength $n$, and $\bar \alpha=1-\alpha$. 
\end{definition}

\begin{figure}[htbp]\vspace{-4mm}
  \centering
  \includegraphics[width=0.4\textwidth]{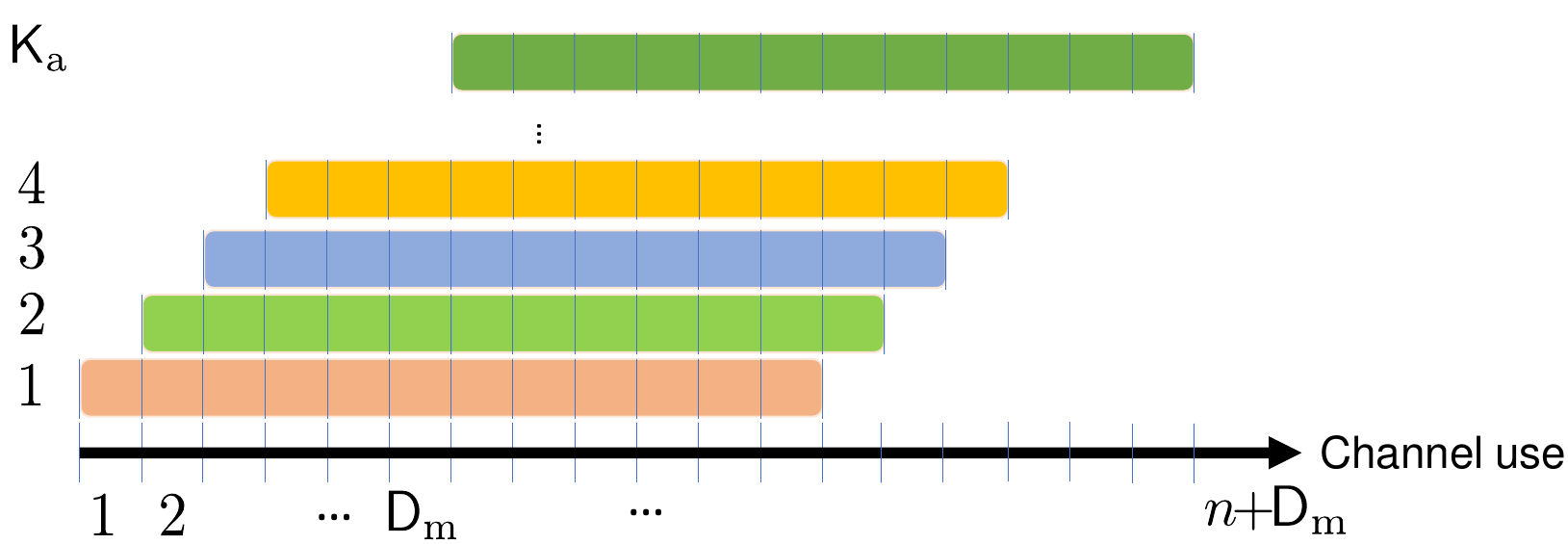}\vspace{-1mm}
  \caption{A $\ka$-active-user AUMAC with $\setd\!=\![0,\!1,\!2,\!3,\!...,\!\dm]$. 
  }\vspace{-1mm}
  \label{fig: AUMAC example}
\end{figure}\vspace{-2mm}

For any $\ell\in[n]$, we define a shift function $\tau_{d_i}(X^n_{i},\ell):=X_{i,\ell-d_i}$, where $X_{i,\ell-d_i}$ is the $(\ell-d_i)$-th element of $X^n_i$, and if $\ell-d_i\not \in[n]$, $X_{i, \ell-d_i}=0,\ \forall i\in[\ka]$. 
The received symbol at the receiver at time $\ell\in [n+\dm]$ is
\begin{align}
	 {Y}_\ell = \sum_{i=1}^{\ka }\tau_{d_i}\left(X^n_i,\ell\right)+  Z_\ell, \label{eq:system_model2}
\end{align}
where the channel input $X^n_i \!\in \!\mathcal X ^n\!\subset\! \mathbb R^n$, 
% \begin{align}
    $\mathcal X^n\!:=\!\{x^n \!:\! x^n\! \in \!\mathbb R^n,\; \|x^n\|^2\!\leq\! n\pow'\}$
% \end{align}
satisfying the maximal power constraint, and $Z_\ell\!\sim\! \mathcal N(0, 1)$, $\forall \ell\!\in\![n\!+\!\dm]$. The received signal of a $\ka$-active-user AUMAC with $\setd\!=\![0,\;1,\;2,\;3,\;...,\;\dm]$ is illustrated in Fig. \ref{fig: AUMAC example}.

We define an AUMAC code, where the receiver decodes without perfect delay information, as follows:
\begin{definition}\label{definition CD2}
    An $(n,\textsfM, \epsilon, \ka, \alpha)-$code for an AUMAC described by $P_{Y|X_{[\ka]}}$ consists of one message set $\mathcal M$, one encoder $f:\; \mathcal M \to \mathcal X^n$, and one decoder $\hat g:\; \mathbb R^{n+\dm}\to \binom{\mathcal M}{\ka}$ such that for the maximal power constraint $\pow'$ and any $\setd$ satisfying $d_i \!\leq\! \alpha n,\, \forall i\!\in\! [\ka]$, the PUPE satisfies
    \begin{align}
    {\mbox{P}_{\text{PUPE}}}:=\sum_{i=1}^{\ka}\frac{1}{\ka}\mbox {Pr}(\tilde {\mathcal E}_i|\setd)\leq \epsilon, \label{definition PUPE constraint}
    \end{align} 
   where $\binom{\mathcal M}{\ka}\vspace{0.8mm}$ is a set
containing $\ka$ distinct elements from the set $\mathcal M$, $\tilde{\mathcal E}_i:=  \{ \{M_i  = M_\ell,\, \forall \ell\in [\ka],\, \ell\neq i \} \cup \{M_i \not\in  \hat g({Y}^{n+\dm}) \} \cup \{\|f(M_i)\|^2 >  n\pow'\} \}$,  $i   \in   [\ka]$, and $M_i$ is the $i$-th transmitted message, which is uniformly distributed in $\mathcal M$.
\end{definition}

For the AUMAC with perfect delay information at the receiver, we define an AUMAC code as follows:
\begin{definition}\label{definition CD3}
    An $(n,\textsfM, \epsilon, \ka, \alpha, \setd)-$code for an AUMAC described by $P_{Y|X_{[\ka]}}$ consists of one message set $\mathcal M$, one encoder $f:\;\mathcal M\to \mathcal X^n$, and one decoder $\hat g:\; \mathbb R^{n+\dm}\to\binom{\mathcal M}{\ka}$ 
    such that \eqref{definition PUPE constraint} is satisfied for the maximal power constraint $\pow'$ and given $\setd$ satisfying $d_i \leq \alpha n,\, \forall i\in [\ka]$.
\end{definition}

%% file: Contents/Main_result_arXiv.tex
% !TeX root = ../ISIT2024.tex

\section{Main Results}\vspace{-0mm}
This section investigates the achievable PUPE upper bound of the AUMAC where the receiver decodes based on $n+\dm$ received symbols. In Subsection \ref{subsection, Decoding Without Delay Information}, Theorem \ref{thm CD2} provides the upper bound of the PUPE for the AUMAC without delay information at the receiver. In Subsection \ref{subsection, subsection, Decoding With Perfect Delay Information}, Theorem \ref{thm CD with delay information} provides the upper bound of the PUPE for the AUMAC with delay information at the receiver.

\subsection{Decoding Without Delay Information} \label{subsection, Decoding Without Delay Information}

To analyze the PUPE of the AUMAC with the receiver decoding with $n+\dm$ received symbols, we consider the wrap-decoder for the AUMAC without delay information defined as follows: 
\begin{align}
        \hat g(\tilde {Y}^n)= \argmin{X^n_{[\ka]}\in \mathcal C,\; d_{[\ka]}\in[0,\dm]^{\ka}}\ \Big\|\tilde {Y}^n-X^n_{[\ka],d_{[\ka]}}\Big\|^2,\label{def: cyclic decoder}\vspace{-5mm}
\end{align}
where $\mathcal C $ contains $\textsfM$ codewords,  $X^n_{[\ka],d_{[\ka]}}:=\sum_{i\in [\ka]}\matr{S}^{d_i}X^n_i$, $\matr{S}:=\big[\begin{smallmatrix}
  0^{n-1} & 1\\
  \matr{I}_{n-1} & (0^{n-1})^\top
\end{smallmatrix}\big]$ 
is the circular shift matrix, $\matr{I}_{n-1}$ is an $(n-1) \!\times\! (n-1)$ identity matrix, $\tilde {Y}^n\!:=\![Y_1,\, Y_2,\,...,\, Y_{n}]+[Y_{n+1},\, Y_{n+2},\,...,\, Y_{n+\dm},0^{n-\dm}]$, and $0^{k}$ is a vector of zeros of length $k$. Because the receiver does not have delay information, the decoder compares all combinations of delays $d_{[\ka]}\!\!\in\!\![0,\dm]^{\ka}\!$ and codewords $\!X^n_{[\ka]}\!\!\in\! \mathcal C$. 

Based on Definition \ref{definition CD2} and the decoder defined in \eqref{def: cyclic decoder}, we summarize the FBL analysis results for the AUMAC as follows:
\begin{theorem}
    \label{thm CD2}
Fix $0< \pow \leq \pow'$. There exists an $(n,\textsfM, \epsilon, \ka,\alpha)-$code for an AUMAC such that the PUPE is upper-bounded as follows.
    \begin{multline}
    \sum_{s=1}^{\ka}\frac{s}{\ka}\binom{\ka}{s}\bigg\{\!
    \min_{t>0}\bigg\{
    \frac{\exp\left(g(s,t,T_s)\right)}
    {T_s(1\!-\!T_s)\sqrt{2\pi}}\left(g^{(2)}_{{t_s}}(s,t,T_s)\!\right)^{\frac{-1}{2}}\bigg\}\\
    \!\!+\!
    \sum_{s_1=1}^{\ka\!-\!s} \binom {\ka\!\!-\!\!s}{s_1 }(1+\alpha n)^{s_1}\min_{t>0}\bigg\{\frac{\exp\left(\bar g(s,s_1, t,\underline T_s)\right)}
    {T^*\sqrt{2\pi}}\\
    \cdot \left(g^{(2)}_{{t_s}}(s, t,\tilde {T}_s)\right)^{\frac{-1}{2}}\bigg\}\bigg\}
    \!+\!p_0\!+\!O\left(\frac{\exp(-n)}{\sqrt{n}}\right)\!\leq\! \epsilon,\label{eq: thm cd2 main result}\vspace{-0.5mm}
\end{multline}
where $T_s\in\left(0,\min\left\{\frac{1}{4t},1\right\}\right)$ such that $g^{(1)}_{{t_s}}(s,t,T_s)=0,$ $\underline T_s\in\left(0,\min\left\{\frac{1}{4t},1\right\}\right)$ such that 
\begin{align}
     g^{(1)}_{{t_s}}(s,t,\underline T_s)=\underline \gamma(s,s_1, t,\underline T_s),\label{eq: the condition of lower bound of saddlepoint}\vspace{-1mm}
\end{align}
$\bar T_s\in\left(0,\min\left\{\frac{1}{4t},1\right\}\right)$ such that
\begin{align}
        g^{(1)}_{{t_s}}(s,t,\bar T_s)=\bar \gamma(s,s_1,t,\bar T_s),\label{eq: the condition of upper bound of saddlepoint}
    \end{align}
\begin{flalign}
    g(s,t,t_s):=& t_s\!\log\theta(s)\!-\!\frac{n(t_s\!\!-\!1)}{2}\!\log (1\!+\!\!2s\pow t)\nonumber\\
    &-\!\frac{\alpha n }{2}\!\log (1\!+\!2s\pow t(1\!+\!t_s)(1\!-\!4 t\!\cdot\! t_s))\nonumber\\
    &-\frac{\bar \alpha n }{2}\log \left(1\!+\!2s\pow t(1\!+\!t_s)(1\!-\!2 t\cdot t_s)\right),\label{eq: theorem, without delay info}\\
    \theta(s)=&\binom{\textsfM-\ka}{s}(1+\alpha n )^s,\\
    \bar g(s,s_1,t,t_s):=&g(s,t,t_s)- \frac{2 n s_1 \pow (\alpha f_2+\bar \alpha f_1)}{\sqrt{1+2\bar \lambda+\frac{\bar \lambda^2}{3}}},\\
    T^*:=&\min\{\bar T_s(1-\bar T_s),\; \underline T_s (1-\underline T_s)\},\label{eq:def of T star}\\
    \tilde T_s:=&\arginf {t_s\in (0, \min\{1,1/4t\})} g^{(2)}_{t_s}(s,t,t_s),\\
    % \bar \lambda :=&4s_1 f_1 \pow\\
    f_\iota(s,t,t_s):=&\frac{(1-2\cdot \iota \cdot t\cdot t_s)t\cdot t_s}{1\!+\!2s\pow t(1\!+\!t_s)(1\!-\!2\cdot\iota  \cdot t \cdot t_s)},\ \iota=\{1,2\},\nonumber\\
    % \tilde T_s:=&\arginf {t_s\in (0, \min\{1,1/4t\})} g_b(s,s_1, t,t_s),\\
    \bar \gamma(s,s_1, t,t_s)\!:=&
    \begin{cases}
        2s_1 n \pow\left( \alpha \frac{\partial f_2}{\partial t_s}\!+\!\bar \alpha \frac{\partial f_1}{\partial t_s}\right),&\text{if } \frac{\partial f_2}{\partial t_s}\geq 0\\
        4s_1 n \pow \frac{\partial f_1}{\partial t_s},&\text{o.w., } \\
    \end{cases}\label{eq: the condition of the bar gamma }\\
    % &\hspace{-6.2em}\text{and}\nonumber\\
    \underline \gamma(s,s_1, t,t_s):=&
    \begin{cases}
        \dfrac{2s_1 n\pow}{1\!+\!2\bar\lambda} \left( \alpha \frac{\partial f_2}{\partial t_s}\!+\!\bar \alpha \frac{\partial f_1}{\partial t_s}\right),&\text{if } \frac{\partial f_2}{\partial t_s}\geq 0\\
        4s_1 n \pow \frac{\partial f_2}{\partial t_s},&\text{o.w., }\!\!\!\\
    \end{cases}\label{eq: the condition of underline gamma}
\end{flalign}
$\bar \lambda \!:=\!4s_1 f_1 \pow$, and $p_0\!:=\!\frac{\ka(\ka\!-\!1)}{2\textsfM}\!+\!\sum\limits_{i=1}^{\ka }\mbox{Pr}(\|X^n_i\|^2>n\pow')$. 
\end{theorem} 

The proof is relegated to Appendix \ref{appen: proof of theorem CD2}.

For an AUMAC without delay information at the receiver, the event that the receiver incorrectly decodes the messages transmitted by the user in $[s]$ consists of the following cases: \vspace{-1mm}
\begin{enumerate}
    \item The codewords transmitted from the users in $[s]$ are incorrectly decoded with any $\hat d_{[s]}\in\{0,\dm\}^s$ and the codewords transmitted from $[\ka]\setminus [s]$ are correctly decoded with correct delays $d_{[\ka]\setminus [s]}$.
    \item The codewords transmitted from the users in $[s]$ are incorrectly decoded with any $\hat d_{[s]}\in\{0,\dm\}^s$ and the codewords transmitted from $[\ka]\setminus [s]$ are correctly decoded with $s_1$ incorrect delays and $\ka-s-s_1 $ correct delays, where $s_1\in[\ka-s]$. 
\end{enumerate}\vspace{-1mm}
Therefore, by utilizing the union bound, we express the probability that the receiver incorrectly decodes the messages transmitted by the users in $[s]$ as the sum of the probabilities of the first case and the second case in \eqref{eq: thm cd2 main result}.  
It is worth noting that the wrap-decoder, AWGN, and the i.i.d. Gaussian codebooks cause the PUPE to be identical for any delay $\setd\in\{0,\dm\}^{\ka}$ because any $\setd\in\{0,\dm\}^{\ka}$ has the same distribution of $\tilde Y$. 
As a result, Theorem \ref{thm CD2} is independent of $\setd$.

\subsection{Decoding With Perfect Delay Information} \label{subsection, Decoding With Perfect Delay Information}

To analyze the PUPE of the AUMAC without delay information, we consider the wrap-decoder defined as follows:
\begin{align}
        \hat g(\tilde {Y}^n)= \argmin{X^n_{[\ka]}\in \mathcal C,\; d_{[\ka]}\in\mathfrak{S}(\setd)}\ \Big\|\tilde {Y}^n-X^n_{[\ka],d_{[\ka]}}\Big\|^2, \label{def: cyclic decoder with delay information}
\end{align}
where $\mathfrak{S}(\setd)$ contains all permutations of $\setd$. Compared to \eqref{def: cyclic decoder}, for AUMAC with perfect delay information at the receiver, we only consider all permutations of $\setd$ instead of all combinations of possible delays. 

 \begin{theorem}
    \label{thm CD with delay information}
Fix $0< \pow \leq \pow'$. There exists an $(n,\textsfM, \epsilon, \ka,\alpha,\setd)-$code for an AUMAC such that the PUPE can be upper-bounded as follows:
    \begin{multline}
    \hspace{-2mm}\sum_{s=1}^{\ka}\frac{s}{\ka}\binom{\ka}{s}\bigg\{\!
    \min_{t>0}\bigg\{
    \frac{\exp\left(g_2(s,t,T_s)\right)}
    {T_s(1\!-\!T_s)\sqrt{2\pi}}\left(g^{(2)}_{2,{t_s}}(s,t,T_s)\!\right)^{-\frac{1}{2}}\bigg\}\\
    \!\!+\!
    \sum_{s_1=1}^{\ka\!-\!s} \binom {\ka\!\!-\!\!s}{s_1 }\frac{(s+s_1)!}{s!}\min_{t>0}\bigg\{\frac{\exp\left(\bar g_2(s,s_1, t,\underline T_s)\right)}
    {T^*\sqrt{2\pi}}\\
    \cdot (g^{(2)}_{2,{t_s}}(s, t,\tilde T_s))^{-\frac{1}{2}}\bigg\}\bigg\}
    \!+\!p_0\!+\!O\left(\frac{\exp(-n)}{\sqrt{n}}\right)\!\leq\! \epsilon,
\end{multline}
where $T_s\in\left(0,\min\left\{\frac{1}{4t},1\right\}\right)$ such that $g^{(1)}_{2,{t_s}}(s,t,T_s)=0,$ $\underline T_s\in\left(0,\min\left\{\frac{1}{4t},1\right\}\right)$ such that 
     $g^{(1)}_{2,{t_s}}(s,t,\underline T_s)=\underline \gamma(s,s_1, t,\underline T_s)$,
$\bar T_s\in\left(0,\min\left\{\frac{1}{4t},1\right\}\right)$ such that
     $g^{(1)}_{2,{t_s}}(s,t,\bar T_s)=\bar \gamma(s,s_1,t,\bar T_s)=0$,   
\begin{flalign}
    g_2(s,t,t_s):=&g(s,t,t_s)-t_s\log\frac{(1+\alpha n)^s}{s!},\label{eq: theorem, with delay info}\\
    \bar g_2(s,s_1,t,t_s):=&g_2(s,t,t_s)- \frac{2 n s_1 \pow (\alpha f_2+\bar \alpha f_1)}{\sqrt{1+2\bar \lambda+\frac{\bar \lambda^2}{3}}},
\end{flalign}
the terms $g(s,t,t_s)$, $\bar \lambda $, $T^*$, $\tilde T_s$, $ f_1(s,t,t_s)$, $ f_2(s,t,t_s)$, $\bar \gamma(s,s_1, t,t_s)$, and $\underline \gamma(s,s_1, t,t_s)$ are defined in Theorem \ref{thm CD2}. 
\end{theorem} 

The proof is relegated to Appendix \ref{appen: proof of theorem with delay information}.

Note that for a given number of incorrectly decoded codewords, $s\in[\ka]$, we derive $\bar \gamma(s,s_1, t,t_s)$ and $\underline \gamma(s,s_1, t,t_s)$ by finding the upper bound for a given number of the incorrect delays paired to the correctly decoded codewords, $s_1\in[0, \ka\!-\!s]$. Even though the delay information is available at the receiver, the receiver may correctly decode $\ka-s$ codewords with $s_1$ incorrect delays. As a result, $\bar \gamma(s,s_1, t,t_s)$ and $\underline \gamma(s,s_1, t,t_s)$ are identical regardless of the availability of delay information. 

% If the receiver has perfect information on delay, for a given $s$, there are $s!$ permutations of the delays for the incorrectly decoded codewords. However, if the receiver does not have perfect information on delay, there are $(1+\alpha n )^s$ combinations of the delays for the incorrectly decoded codewords. These two facts result in \eqref{eq: theorem, with delay info}. The proof is relegated to Appendix \ref{appen: proof of theorem with delay information}.

%% file: Contents/Numerical.tex
% !TeX root = ../ISIT2024.tex

\section{Numerical Results}\vspace{-0.5mm}
Based on the results of Theorem \ref{thm CD2} and Theorem \ref{thm CD with delay information} without considering $O\left(\frac{\exp(-n)}{\sqrt{n}}\right)$, we numerically evaluate $\ebno$ versus $\ka$, where $\ebno:=\frac{n\pow'}{2\log_2 \textsfM}\vspace{0.1em}$. 
The PUPE upper bounds from Theorem \ref{thm CD2} and Theorem \ref{thm CD with delay information} are compared to UMAC and different AUMAC schemes under different scenarios with the following parameters: $\logM=100$, $n=4000$, $\epsilon=5\times 10^{-2}$, and $\ka\in[50, 160]$. In Figure \ref{fig: AUMAC result}, all curves are evaluated by numerically optimizing $\pow<\pow'$.
The black solid curve indicates the required $\ebno$ of the UMAC \cite{Polyanskiy_perspective_on_UMAC}. A blue dashed curve marked with triangles indicates the required $\ebno$ of the AUMAC where the receiver decodes at the $n$-th channel use with perfect delay information~\cite{ISIT_worst_case}.

We evaluate the required $\ebno$ of the AUMAC without delay information by Theorem \ref{thm CD2} for  $\alpha=0.2$ and $\alpha=0.4$ and that of the AUMAC with delay information by Theorem \ref{thm CD with delay information} with $\alpha=0.2$.
Numerical results show that for an AUMAC, a larger $\alpha$ causes the transmitters to consume more energy to transmit reliably. Observing the curves of Theorem \ref{thm CD2} and Theorem \ref{thm CD with delay information} with $\alpha =0.2$, we conclude that the availability of delay information at the receiver only slightly influences the required $\ebno$. By comparing the $\ebno$ of the AUMAC with delay information evaluated by Theorem \ref{thm CD with delay information} to the $\ebno$ of \cite{ISIT_worst_case}, the energy efficiency is improved significantly by decoding with completely received codewords, especially when $\ka$ is large.

\begin{figure}
\centering
\begin{tikzpicture}

\begin{axis}[
    xlabel={The number of active users ($\ka$)},
    ylabel={$E_b/N_0$, dB},
    xmin=50, xmax=160,
    ymin=5, ymax=27,
    xtick={60,80,100,120,140,160},
    ytick={5,10,15,20,25,30},
    legend pos=south east,
    ymajorgrids=true,
    xmajorgrids=true,
    grid style=dashed,
    legend cell align={left},
    legend style={nodes={scale=0.6, transform shape}}
]

\addplot[
    color=cyan,dashed, mark=triangle*,mark options={scale=1.1}
    ]
    coordinates {
        (50,8.61855456709120)	(55,9.61855456709120)	(60,10.6185545670912)	(65,11.6185545670912)	(70,12.7185545670912)	(75,13.8185545670912)	(80,14.8185545670912)	(85,15.9185545670912)	(90,17.1185545670912)	(95,18.2185545670912)	(100,19.3185545670912)	(105,20.5185545670912)	(110,21.6185545670912)	(115,22.8185545670912)	(120,24.0185545670912)	(125,25.2185545670912)	(130,26.3185545670912)	(135,27.5185545670912)	(140,28.7185545670912)	(145,29.9185545670912)	(150,31.2185545670912)	(155,32.4185545670912)	(160,33.6185545670912)
        };
    \addlegendentry{Async. \cite{ISIT_worst_case}, $\alpha =0.2$}

\addplot[
    color=orange,
    dashed,
    mark=*,mark options={scale=1}
    ]
    coordinates {
    (160,27.6385545670912)	(155,26.6785545670912)	(150,25.6985545670912)	(145,24.7385545670912)	(140,23.7585545670912)	(135,22.8185545670912)	(130,21.9185545670912)	(125,20.9185545670912)	(120,20.0185545670912)	(115,19.1185545670912)	(110,18.2185545670912)	(105,17.3185545670912)	(100,16.4185545670912)	(95,15.6185545670912)	(90,14.7185545670912)	(85,13.9185545670912)	(80,13.0185545670912)	(75,12.2185545670912)	(70,11.4185545670912)	(65,10.6185545670912)	(60,9.81855456709120)	(55,8.97855456709120)	(50,8.17855456709120)
    };
    \addlegendentry{Theorem \ref{thm CD2}, $\alpha =0.4$}

    \addplot[
    color=cyan,dashed, mark=*,mark options={scale=1}
    ]
    coordinates {
    (160,26.7185545670912)	(155,25.7585545670912)	  (150,24.7785545670912)	(145,23.7985545670912)	    (140,22.8785545670912)	(135,21.8985545670912)	  (130,20.9585545670912)	(125,20.0185545670912)	    (120,19.0785545670912)	(115,18.1785545670912)	  (110,17.2385545670912)	(105,16.3385545670912)	    (100,15.4385545670912)	(95,14.5185545670912)	  (90,13.6585545670912)	(85,12.8185545670912)	    (80,11.9185545670912)	(75,11.1185545670912)	  (70,10.3185545670912)	(65,9.51855456709120)	    (60,8.71855456709120)	(55,7.87855456709120)	  (50,7.17855456709120)
    };
    \addlegendentry{Theorem \ref{thm CD2}, $\alpha =0.2$}

    \addplot[
    color=purple, mark=square,mark options={scale=1.05}
    ]
    coordinates {
        (160,26.6585545670912)	(155,25.6785545670912)	(150,24.7185545670912)	(145,23.7785545670912)	(140,22.7985545670912)	(135,21.8385545670912)	(130,20.8985545670912)	(125,19.9585545670912)	(120,19.0185545670912)	(115,18.1185545670912)	(110,17.1785545670912)	(105,16.2785545670912)	(100,15.3785545670912)	(95,14.4785545670912)	(90,13.5585545670912)	(85,12.6985545670912)	(80,11.8185545670912)	(75,10.9785545670912)	(70,10.1385545670912)	(65,9.29855456709120)	(60,8.49855456709120)	(55,7.69855456709120)	(50,6.89855456709120)
        };
    \addlegendentry{Theorem \ref{thm CD with delay information}, $\alpha =0.2$}

    \addplot[
    color=black, no marks,
    ]
    coordinates {
        (50,5.51855456709120)	(60,6.91855456709120)	(70,8.41855456709120)	(80,10.0185545670912)	(90,11.6185545670912)	(100,13.2185545670912)	(110,14.9185545670912)	(120,16.6185545670912)	(130,18.3185545670912)	(140,20.0185545670912)	(150,21.8185545670912)	(160,23.6185545670912)      
        };
    \addlegendentry{Sync. UMAC \cite{Polyanskiy_perspective_on_UMAC}}

    %     \addplot[
    % color=black, no marks,
    % ]
    % coordinates {
    % (50,3.47983492387358)	(60,4.37983492387358)	(70,5.37983492387358)	(80,6.37983492387358)	(90,7.47983492387358)	(100,8.57983492387358)	(110,9.67983492387358)	(120,10.7798349238736)	(130,11.9798349238736)	(140,13.0798349238736)	(150,14.2798349238736)	(160,15.4798349238736)
    % };
    % \addlegendentry{UMAC \cite{Polyanskiy_perspective_on_UMAC}, n=5600}
    
\end{axis}
\end{tikzpicture}\vspace{-2mm}
\caption{The trade-off between $\ebno$ and $\ka$}\vspace{-7mm}
\label{fig: AUMAC result}
\end{figure}

%% file: Contents/Conclusion.tex
% !TeX root = ../ISIT2024.tex

\section{Conclusions}\vspace{-1mm}

This work investigates the FBL performance of the AUMAC where the receiver decodes with $n+\dm$ channel uses and does not have delay information. Although the asynchronous model and decoding without delay information lead to a more complex PUPE bound, we design a wrap-decoder achieving the same PUPE for all delays. In addition, the wrap-decoder enables us to derive a uniform upper bound of the PUPE for all combinations of codewords and delays. As a result, the analysis is significantly simplified by multiplying the uniform upper bound with the number of the corresponding error events instead of calculating the tail probabilities of all error events. The numerical results show the trade-off between $\ebno$ and delay constraint $\dm$. By observing the $\ebno$, the performance loss due to the asynchronicity is improved by applying the wrap-decoding to decode with completely received codewords. In addition, the wrap-decoder can efficiently overcome the lack of information on delays since the energy efficiencies of the AUMAC with and without delay information are nearly identical.

%% file: back/Appendix_Proof_thm1.tex
% !TeX root = ../ISIT2024.tex

\section{Proof of Theorem~\ref{thm CD2}} 
\label{appen: proof of theorem CD2}
% \subsection{Achievable Result}
% Theorem \ref{thm CD2} is deriv
% the design of the decoder defined in Def. \ref{definition CD2} and \eqref{def: cyclic decoder}.

In the following, we apply the RCU bound \cite{Polyanskiy.2010} to express the PUPE defined in Definition~\ref{definition CD2} as a sum of tail probabilities. Then, we apply the Taylor expansion, the Chernoff bound, and the inverse Laplace transform to the tail probabilities.
% Inspired by \cite{Cyclic_decoding}, for an $(n,\textsfM, \epsilon, \ka,\alpha)-$code and the receiver without perfect delay information, we consider the decoder defined by
% \begin{align}
%         \hat g(\tilde {Y}^n)= \argmin{X^n_{[\ka]}\in \mathcal C, d_{[\ka]}\in[0,\dm]^{\ka}}\ \Big\|\tilde {Y}^n-X^n_{[\ka],d_{[\ka]}}\Big\|^2,\label{def: cyclic decoder}
% \end{align}
% where $X^n_{[\ka],d_{[\ka]}}:=\sum_{i\in [\ka]}\matr{S}^{d_i}X^n_i$, $\matr{S}:=\big[\begin{smallmatrix}
%   0^{n-1} & 1\\
%   \matr{I}_{n-1} & (0^{n-1})^\top
% \end{smallmatrix}\big]$ 
% is the circular shift matrix, $\matr{I}_{n-1}$ is an $(n-1) \!\times\! (n-1)$ identity matrix, $\tilde {Y}^n\!:=\![Y_1,\,Y_2,\,...,\,Y_{n}]+[Y_{n+1},\,Y_{n+2},\,...,\,Y_{n+\alpha n},0^{n-\alpha n}]$, and $0^{k}$ is a vector of \tilde Zeros of length $k$. 
% \begin{align}
%     \text{Pr}(V>U)= \frac{\exp(\log(\mathds{E}[\exp(T_s\log(V))]))}{T_s(1-T_s)\sqrt{2\pi}}\left(\frac{d^2}{d t_s^2}\log(\mathds{E}[\exp(t_s\log(V))])\big|_{t_s=T_s}\right)^{-\frac{1}{2}}+O(\exp(n)/\sqrt{n}),\label{eq:SPA}
% \end{align}
% to the RCU bound to approximate the tail probability, where $T_s$ is the saddlepoint. 
% \subsection{ Per-User Error Probability }

We define ${\mathcal E}: =\{\{M_\ell \neq M_i,\;\forall i \neq \ell,\; \forall i,\ell \in [\ka]\}\cap \{\|f(M_i)\|^2\leq n\pow',\;\forall i \in [\ka]\}\}$, which represents the event that all transmitted messages are distinct and transmitted codewords fulfill the power constraint.
By Definition~\ref{definition CD2}, for any $\setd\in[0,\dm]^{\ka}$, the PUPE of an $(n, \textsfM, \epsilon, \ka, \alpha)$-code can be upper-bounded as follows:
\begin{flalign*}
    {\mbox{P}_{\text{PUPE}}}
    &\!\leq\! p_0\!+\!\!\sum_{\sett\subseteq [\ka]}\!\frac{|\sett|}{\ka}\mbox{Pr}\big( M_\sett\!\not\in\!\hat  g(\tilde Y^n), M_{\bar \sett}\!\in\! \hat  g(\tilde Y^n)|\setd, {\mathcal E}\Big),
\end{flalign*}
where $\bar \sett\!:=\![\ka]\setminus\sett$ and $p_0$ is defined in Theorem~\ref{thm CD2}. 
% The decoder defined in \eqref{def: cyclic decoder}, i.i.d. codewords, and AWGN result in identical PUPE for all $\setd\!\in\![0,\dm]^{\ka}$.
% To simplify the notations, we omit $\setd$ and $\mathcal E$. 
By substituting the wrap-decoder defined in \eqref{def: cyclic decoder}, we have
\begin{flalign}
    &\!\!\sum_{\sett\subseteq [\ka]}\frac{|\sett|}{\ka}\mbox{Pr}\big( M_\sett\!\not\in\!\hat  g(\tilde Y^n), M_{\bar \sett}\!\in\! \hat  g(\tilde Y^n)|\setd, {\mathcal E}\Big)\nonumber\\
    &\!\!=\!\sum_{s=1}^{\ka}\!\frac{s\binom{\ka}{s}}{\ka} \!\mbox{Pr}\!\left( \raisebox{1.6ex}{$\displaystyle 
    \!\!\bigcup_{\substack{{\bar  {X}^n_{[s]}\!\subseteq \!f\!(\!\mathcal M\setminus M_{[\ka]}\!),}\\{\hat d_{[\ka]}\in[0,\dm]^{\ka}}}}$}\Big\|\tilde Y \!\!- \!\!\bar {X}^n_{[s],\hat d_{[s]}} \!\!\!- \!\!X^n_{\bar{[s]},\hat d_{\bar{[ s]}}}\!\Big\|^2 \!\leq  \!\|\tilde Z^n\|^2\!\! \right)\!, 
    \label{eq: initial error event}
\end{flalign}
where $\bar {[s]}:= [\ka]\setminus [s]$, $\tilde Z_{i} \sim \mathcal N(0,2),\; i\in[\dm]$, and $\tilde Z_{i} \sim \mathcal N(0,1),\; i\in[\dm]$ due to the wrap-decoding. The equality is because, for the wrap-decoder defined in \eqref{def: cyclic decoder}, $\tilde Y \in \mathbb R^n$ is the sum of $\ka$ i.i.d. codewords plus independent noise. Therefore, all subsets $\sett\!\subseteq\![\ka]$ with the same cardinality $|\sett|$ lead to the same tail probability. 

For the AUMAC without perfect delay information at the receiver, the receiver must consider all possible delays, i.e., all $\hat d_{[\ka]}\!\in\![0,\dm]^{\ka}$.
Then, the RCU bound \cite{Polyanskiy.2010} and the Chernoff bound with $t>0$ are applied to \eqref{eq: initial error event} as follows:
\begin{flalign}
    & \mbox{Pr}\left( \raisebox{1.6ex}{$\displaystyle\bigcup _{\substack{{\bar {X}_{[s]}\subseteq f(\mathcal M\setminus M_{[\ka]}),}\\{\hat d_{[\ka]}\in[0,\dm]^{\ka}}}}$}\Big\|\tilde Y^n-\bar {X}^n_{[s],\hat d_{[s]}}-X^n_{\bar {[s]},\hat d_{\bar {[s]}}}\Big\|^2\leq \|\tilde Z^n\|^2\right)\nonumber\\
    &=\mathds{E} \left[ \mbox{Pr}\left( \raisebox{1.6ex}{$\displaystyle\bigcup _{\substack{{\bar {X}_{[s]}\subseteq f(\mathcal M\setminus M_{[\ka]}),}\\{\hat d_{[\ka]}\in[0,\dm]^{\ka}}}}$}\Big\|\tilde Y^n-\bar {X}^n_{[s],\hat d_{[s]}}-X^n_{\bar {[s]},\hat d_{\bar {[s]}}}\Big\|^2\right.\right.\nonumber\\
    &\hspace{45mm} \left.\left.\leq \|\tilde Z^n\|^2\bigg| X^n_{[\ka]},\tilde Z^n\right)\right]\nonumber\\
    &\leq \mathds{E} \left[ \min\left\{1,\sum_{\hat d_{[\ka]}\in[0,\dm]^{\ka}}\binom{\textsfM-\ka}{s}\right.\right.\nonumber\\
    &\quad \mbox{Pr}\left( \Big\|\tilde Y^n-\bar {X}^n_{[s],\hat d_{[s]}}-X^n_{\bar {[s]},\hat d_{\bar {[s]}}}\Big\|^2\left.\left.\leq \|\tilde Z^n\|^2\bigg| X^n_{[\ka]},\tilde Z^n\right)\right]\right\}\nonumber\\
    &\leq \sum_{\substack{\hat d_{\bar {[s]}}\in[0,\dm]^{\ka-s}\setminus d_{\bar{[s]}}}}
    \min_{t>0}\mathds{E}\bigg[\min\bigg\{1,\theta(s)  \exp(t\|\tilde Z^n\|^2)\nonumber\\
    &\hspace{0.5cm} \cdot \mathds{E}\bigg[\exp\bigg(-t\Big\|B^n_{\bar {[s]}}+X^n_{[s],d_{[s]}}-\bar {X}^n_{[s],\hat d_{[s]}}+\tilde Z^n\Big\|^2\bigg)\bigg]
    \bigg\}\bigg]\!\!\!\nonumber\\
    &\quad +\min_{t>0}\mathds{E}\bigg[\min\bigg\{1,\theta(s)  \exp(t\|\tilde Z^n\|^2)\nonumber\\
    & \hspace{0.8cm}\cdot\mathds{E}\bigg[\exp\bigg(-t\Big\|X^n_{[s],d_{[s]}}-\bar {X}^n_{[s],\hat d_{[s]}}+\tilde Z^n\Big\|^2\bigg)\bigg]\bigg\}\bigg],\!\!\!
    \label{eq: Chernoff bound; RCU bound}
\end{flalign}
where $\theta(s):=\binom{\textsfM-\ka}{s}(1+\dm)^s$, $B^n_{\bar{[ s]}}:=X^n_{\bar{[ s]},d_{\bar{[s]}}}-X^n_{\bar{[ s]},\hat d_{\bar{[ s]}}}$, and $B^n\sim \mathcal N(0^n, \matr{\Sigma}(\hat d_{\bar{[ s]}}))$. 
% The inequality follows from the RCU bound and the Chernoff bound with parameter $t>0$. 
The first equality follows from the random coding \cite{Polyanskiy.2010}. The first inequality follows from the union bound. The last inequality follows from the Chernoff bound and $\min\{1, a+b\}\leq \min\{1, a\}+\min\{1, b\}$ for $a, b\geq 0$.
The first term in \eqref{eq: Chernoff bound; RCU bound} represents the probability that the receiver incorrectly decodes the codewords transmitted by users in $[s]$ and correctly decodes codewords with incorrect delay $\hat d_{\bar{[s]}}$; the second term represents the probability that the receiver incorrectly decodes the codewords transmitted by users in $[s]$ and correctly decodes codewords with correct delay $d_{\bar{[s]}}$.

To derive tail probabilities, the Taylor expansion and the inverse Laplace transform are used in \cite{Saddlepoint_approximation_keystone,saddlepoint_of_RCUs, ISIT_worst_case}.
Therefore, for a given $s\in[\ka]$, $\hat d_{\bar{[ s]}}\in [0, \dm]^{\ka-s}\setminus d_{\bar{[ s]}}$ and $t>0$, we apply the Taylor expansion and the inverse Laplace transform to the first term in \eqref{eq: Chernoff bound; RCU bound} as follows:
\begin{flalign}
   &\min_{t>0}\mathds{E}\bigg[\min\bigg\{1,\theta(s)  \exp(t\|\tilde Z^n\|^2)\nonumber\\
    &\hspace{0.6cm} \cdot \mathds{E}\bigg[\exp\bigg(-t\Big\|B^n_{\bar {[s]}}+X^n_{[s],d_{[s]}}-\bar {X}^n_{[s],\hat d_{[s]}}+\tilde Z^n\Big\|^2\bigg)\bigg]\bigg\}\bigg]\!\!\!\nonumber\\&\leq\frac{\exp\left(g(s,t,T_s)\!\!+\!\log\mathds{E}\!\left[\exp\left(\!-(B_{\bar{[s]}}^n)^\top \!\matr{A}(s,t,t_s) B_{\bar{[s]}}^n\right)\!\right]\!\right)}{T_s(1-T_s)\sqrt{2\pi}}\nonumber\\
    &\qquad  \cdot\! \left(\!\frac{d^2}{d t_s^2}\!\log\mathds{E}\left[\exp\left(\!-\!(B_{\bar{[s]}}^n)^\top \matr{A}(s,t,t_s) B_{\bar{[s]}}^n\right)\right]\bigg|_{t_s=T_s}\right.\nonumber\\
    &\qquad \qquad+g_{t_s}^{(2)}(s,t,T_s)\!\bigg)^{-\frac{1}{2}}+O\bigg(\frac{\exp(-n)}{\sqrt{n}}\bigg),\label{eq: initial SPA with B}
\end{flalign}
where $T_s\in\left(0,\min\left\{\frac{1}{4t},1\right\}\right)$ such that 
\begin{multline}
    \frac{d}{d t_s}\!\log\mathds{E}\left[\exp\left(\!-\!(B_{\bar{[s]}}^n)^\top \matr{A}(s,t,t_s) B_{\bar{[s]}}^n\right)\right]\bigg|_{t_s=T_s}\\
    =-g^{(1)}_{t_s}(s,t,T_s),
\end{multline}
\begin{multline}
    g(s,t,t_s):= t_s\!\log\theta(s)\!-\!\frac{n(t_s\!\!-\!1)}{2}\!\log (1\!+\!\!2s\pow t)\\
    \!-\frac{\alpha n }{2}\!\log (1\!+\!2s\pow t(1\!+\!t_s)(1\!-\!4 t\!\cdot\! t_s))\\
    -\frac{\bar \alpha n }{2}\log \left(1\!+\!2s\pow t(1\!+\!t_s)(1\!-\!2 t\cdot t_s)\right),
\end{multline}
the matrix $\matr{A}(s,t,t_s)\in \mathbb R^{n\times n}$ is a diagonal matrix, $A(s,t,t_s)_{i,i}=f_2(s,t,t_s),\ \forall i\in [\dm]$, and $A(s,t,t_s)_{i,i}=f_1(s,t,t_s),\ \forall i\in [n]\setminus[\dm]$, 
$$f_1(s,t,t_s)=\frac{(1-2t\cdot t_s)t\cdot t_s}{1+2s\pow t(1+t_s)(1-2 t\cdot  t_s)},$$
and
$$f_2(s,t,t_s)=\frac{(1-4t\cdot t_s)t\cdot t_s}{1+2s\pow t(1+t_s)(1-4 t\cdot t_s)}.$$
To simplify the notation, we use $\matr{A},\ f_1,$ and $f_2$ to denote $\matr{A}(s,t,t_s),\, f_1(s,t,t_s)$, and $f_2(s,t,t_s)$, respectively.
It is worth noting that to guarantee the convergence of $\tilde g(s,s_1,t,t_s)$ in \eqref{eq: initial SPA with B} for a given $t\!>\!0$, we need $0\!<\!t_s\!<\!\frac{1}{4t}$, and to guarantee a positive $T_s(1-T_s)$ in \eqref{eq: initial SPA with B}, we need $0\!<\!t_s\!<\!1$. If $\underline T_s$ and $\bar T_s$ satisfy \eqref{eq: the condition of lower bound of saddlepoint} and \eqref{eq: the condition of upper bound of saddlepoint}, respectively, but at least one of them does not belong to the interval $\left(0,\min\left\{\frac{1}{4t},1\right\}\right)$, then the error probability is considered to be $1$.

Let $s_1 :=\sum_{i\in\bar{[s]}}\ind(d_{i}\neq \hat d_i)$ be the number of incorrect delays paired to correctly decoded codewords and $\matr{G}(\hat d_{\bar{[ s]}}) \matr{G}(\hat d_{\bar{[ s]}})^\top=\matr{\Sigma}(\hat d_{\bar{[ s]}})$. We derive the expectation by expressing the quadratic form as follows \cite[Corollary 3.2a.2]{Quadratic_form}:
\begin{flalign}
    \hspace{-2mm}\mathds{E}\left[\exp\left(-(B^n)^\top \matr{A} B^n\right)\right]
    &=\prod_{i=1}^r \bigg(1+2\lambda_i\Big(\matr{\Sigma}\big(\hat d_{\bar{[ s]}}\big)\Big)\bigg)^{-\frac{1}{2}},\label{eq: the result from quadratic form}
\end{flalign}
where 
% $W_{[r]}$ are i.i.d. standard Gaussian random variables, 
$r=\text{rank}\Big(\matr{\Sigma}\big(\hat d_{\bar{[ s]}}\big)\Big)$, and $\lambda_{[r]}\Big(\matr{\Sigma}\big(\hat d_{\bar{[ s]}}\big)\Big)$ are non-zero eigenvalues of $\matr{G}\big(\hat d_{\bar{[ s]}}\big)^\top \matr{A} \matr{G}\big(\hat d_{\bar{[ s]}}\big)$. To simplify the notations, we use $\lambda_{[r]}$ to represent $\lambda_{[r]}\Big(\matr{\Sigma}\big(\hat d_{\bar{[ s]}}\big)\Big)$.
% To simplify the notations, we use $\Sigma,\ G$ and $\lambda_{[r]}$ to represent $\Sigma\big(\hat d_{\bar{[ s]}}\big),\ G\big(\hat d_{\bar{[ s]}}\big)$ and $\lambda_{[r]}\Big(\Sigma\big(\hat d_{\bar{[ s]}}\big)\Big)$, respectively.
% \textit{Proof}: todo
% For a given $t$, $s$, and $\hat d_{\bar{[s]}}$, 
We substitute \eqref{eq: the result from quadratic form} into \eqref{eq: initial SPA with B} as follows:
\begin{align}
   &\mathds{E}\Big[\min\Big\{1,\theta(s)  \exp(t\|\tilde Z^n\|^2)\nonumber\\
    &\hspace{0.85cm} \cdot \mathds{E}\left[\exp\left(-t\|B^n_{\bar {[s]}}+X^n_{[s],d_{[s]}}-\bar {X}^n_{[s],\hat d_{[s]}}+\tilde Z^n\|^2\right)\right]
    \Big\}\Big],\!\!\!\nonumber\\
    &\leq\frac{\exp\left(\tilde g(s,s_1,t,T_s)\right)}
    {T_s(1-T_s)\sqrt{2\pi}}\left( \tilde g^{(2)}_{t_s}(s,s_1,t,T_s)
    \right)^{-\frac{1}{2}}\!+\!O\!\bigg(\!\frac{\exp(\!-n\!)}{\sqrt{n}}\!\bigg),
    \label{eq: CD2 main result, shifted of correct codewords, for a certain incorrect delay pattern}
\end{align}
where $T_s\in\left(0,\min\left\{\frac{1}{4t},1\right\}\right)$ such that $\tilde g_{t_s}^{(1)}(s,s_1,t,T_s)=0$, 
\begin{flalign}
    \tilde g(s,s_1,t,t_s):= &g(s,t,t_s)-\frac{1}{2}\sum_{i=1}^{r} \log (1+2\lambda_i). \label{eq: appendix 1, the condition of T_s}
\end{flalign}
For a give $\hat d_{\bar{[s]}}$, \eqref{eq: CD2 main result, shifted of correct codewords, for a certain incorrect delay pattern} provides the probability that the decoder incorrectly decodes the codewords transmitted by users in $[s]$ and correctly decoded codewords transmitted by user in the set $\bar {[s]}$ with delays $\hat d_{\bar{[s]}}$.
However, the first term in \eqref{eq: Chernoff bound; RCU bound} requires calculating all possible $\hat d_{\bar{[s]}}\in[0,\alpha n]^{(\ka-s)} \setminus d_{\bar{[s]}}$. It requires calculating $(1+\dm)^{(\ka-s)}$ different combinations of the delays, which is infeasible when $\ka$ and $\dm$ are large. To overcome the computational complexity, we derive the uniform upper bound of the probabilities for a given $s_1$ by upper-bounding $\tilde g(s,s_1,t, T_s)$, $(T_s(1-T_s))^{-1}$, and $\left( \tilde g^{(2)}_{t_s}(s,s_1,t, T_s)
    \right)^{-\frac{1}{2}}$ in \eqref{eq: CD2 main result, shifted of correct codewords, for a certain incorrect delay pattern}, separately.

To find a uniform upper bound of the term $(T_s(1-T_s))^{-1}$, we introduce the following lemma. 
\begin{lemma}
    Let $q(t_s)$ be a function with range of convergence $\tilde {\mathcal R}$, $\mathcal R:=\tilde {\mathcal R}\cap (0,1)$, and $q(t_s)$ is convex with respect to $t_s\in \mathcal R$.
    Let $T_s \in \mathcal R $ satisfy $$q^{(1)}(T_s)=\sum_{i=1}^{r}\frac{\frac{\partial }{\partial t_s}\tilde {\lambda}_i\big|_{t_s=T_s}}{1+2\tilde {\lambda}_i},$$ 
    where
    $\tilde {\lambda}_{[r]}$ are non-zero eigenvalues of a matrix $\matr{K}\cdot \matr{A}(t_s)\in \mathbb R^{n \times n}$ with rank $r$, $\matr{K}\in \mathbb R^{n \times n}$ and $\matr{A}(t_s)\in \mathbb R^{n \times n}$ are positive-semidefinite, $\matr{A}(t_s)$ is a diagonal matrix, and $A(t_s)_{i,i}$ is convex with respect to $t_s\in \mathcal R$, $\forall i \in [n]$. Then, there exists $\bar T_s$ and $\underline T_s$ satisfying    
    $q^{(1)}(\bar T_s)=q_1(\bar T_s)$ and 
    $q^{(1)}(\underline T_s)=q_2(\underline T_s),$ respectively, such that $\underline T_s \leq T_s \leq \bar T_s,$ where  
    \begin{align*}
        q_1(t_s):=&
        \begin{cases}
          \mbox{Tr}(\matr{K} \frac{\partial \matr{A}}{\partial t_s}), &\text{if } \underline\varrho  \geq 0\\
          \bar \varrho\cdot \mbox{Tr}(\matr{K }),  & \text{if } \underline \varrho <0 <\bar \varrho\\
           \mbox{Tr}(\matr{K} \frac{\partial \matr{A}}{\partial t_s})/(1+2\Lambda),  & \text{o.w., }
        \end{cases}\\
        q_2(t_s):=&
        \begin{cases}
          \mbox{Tr}(\matr{K} \frac{\partial \matr{A}}{\partial t_s})/(1+2\Lambda), &\text{if } \underline\varrho  \geq 0\\
          \underline \varrho\cdot \mbox{Tr}(\matr{K }),  & \text{if } \underline \varrho <0 <\bar \varrho\\
           \mbox{Tr}(\matr{K} \frac{\partial \matr{A}}{\partial t_s}),  & \text{o.w., }
        \end{cases}
    \end{align*}
    $\Lambda:=\max \{\|\matr{K}\cdot\matr{A}(t_s)\|_1\}\geq \max\{\tilde {\lambda}_{[r]}\}$, $\bar \varrho:= \max_{i\in[n]} \{\frac{\partial }{\partial t_s}A(t_s)_{i,i}\}$, $\underline \varrho:= \min_{i\in[n]} \{\frac{\partial }{\partial t_s}A(t_s)_{i,i}\}$,
    and $\mbox{Tr}(\matr{K})$ represents the trace of the matrix $\matr{K}$.
\end{lemma}

    The proof of Lemma 1 is relegated to Appendix \ref{appendix proof of lemma}.
    
By using Lemma 1, we find $\underline  T_s\in \left(0,\min\left\{\frac{1}{4t},1\right\}\right) $ and $\bar T_s\in \left(0,\min\left\{\frac{1}{4t},1\right\}\right)$ satisfying \eqref{eq: the condition of lower bound of saddlepoint} and \eqref{eq: the condition of upper bound of saddlepoint} such that $\underline T_s\leq T_s\leq \bar T_s$ for a given $\hat d_{\bar{[s]}}$. Therefore, we have the inequality as follows:
\begin{align}
    (T_s(1-T_s))^{-1} \leq (T^*)^{-1}, \label{eq: T^* by using lemma}
\end{align}
where $T^*$ is defined in \eqref{eq:def of T star}. The inequality follows from the concavity of $t_s(1-t_s)$. The terms $\bar T_s$ and $\underline T_s$ are identical for any $\hat d_{\bar{[s]}}$ with a given $s_1$. Therefore, \eqref{eq: T^* by using lemma} is a uniform upper bound for all $\hat d_{\bar{[s]}}$ with a given $s_1$. Additionally, to guarantee that the right-hand side of \eqref{eq: CD2 main result, shifted of correct codewords, for a certain incorrect delay pattern} is positive and converge, we consider
$\underline  T_s,\; \bar T_s \in \left(0,\min\left\{\frac{1}{4t},1\right\}\right)$. Note that, in our case, $f^{(1)}_{1,t_s}(s,t,t_s)\!\geq\! 0$, i.e., $\bar \varrho \!\geq\! 0$, leading to two cases in \eqref{eq: the condition of the bar gamma } and~\eqref{eq: the condition of underline gamma}.

To proceed, we find an upper bound of $\tilde g(s,s_1,t, T_s)$. The non-negative $\frac{\partial ^2}{\partial t_s^2}\tilde g(s,s_1, t,t_s)$ results in the convex $\tilde g(s,s_1, t,t_s)$ with respect to $t_s\in (0, \min\{1,1/4t\})$. Therefore, by substituting $\underline T_s$ into \eqref{eq: appendix 1, the condition of T_s}, we can upper-bound $\tilde g(s,s_1,t,T_s)$ as follows:
\begin{align}
    \tilde g(s,s_1,t,T_s)\leq& g(s,t,\underline T_s)-\frac{1}{2}\sum_{i=1}^{r} \log (1+2\lambda_i)\label{eq: the convexity achieves the min}\\
    \leq& g(s,t,\underline T_s)- \frac{2 n s_1 \pow (\alpha f_2+\bar \alpha f_1)}{\sqrt{1+2\bar \lambda+\frac{\bar \lambda^2}{3}}}\label{eq: applying the definition that sum of eigenvalues is the trace}\\
    :&=\  \bar g(s,s_1,t,\underline T_s), \label{eq: the upper bound of g}
\end{align}
where $T_s$ satisfies $g^{(1)}_{{t_s}}(s,t, T_s)=0$, \eqref{eq: the convexity achieves the min} follows from the fact that the convex function $g(s,s_1,t,t_s)$ achieves its minimum at the point $t_s=T_s$ satisfying $\tilde g^{(1)}_{t_s}(s,s_1, t, T_s)=0$, \eqref{eq: applying the definition that sum of eigenvalues is the trace} follows from inequalities, $\sqrt{1+x+\frac{x^2}{12}}\log(1+x)\geq x$ and $\bar \lambda:=4s_1f_1\pow \geq \max\{\lambda_{[r]}\}$ derived from Gershgorin circle theorem \cite[Theorem 6.1.1]{Horn_Johnson_1985}, and the fact that the sum of all eigenvalues of a matrix is the trace of the matrix. 

To upper-bound $(\tilde g_{t_s}^{(2)}(s,s_1, t,t_s))^{-\frac{1}{2}}$, we find the lower bound of $\tilde g_{t_s}^{(2)}(s,s_1, t,t_s)$ as follows:
\begin{align}
    \tilde g^{(2)}_{t_s}(s,s_1,t,t_s)\!=&g^{(2)}_{t_s}(s,t,t_s)\!-\!\sum_{i=1}^r \frac{\frac{\partial^2}{\partial t_s^2}\lambda_i}{1\!+\!2\lambda_i}\!+\!2 \sum_{i=1}^r \left(\frac{\frac{\partial }{\partial t_s}\lambda_i}{1\!+\!2\lambda_i}\right)^2\\
    \geq & g^{(2)}_{t_s}(s,t,t_s),\label{eq: the negative-semidefinite matrix}
\end{align}
where $\frac{\partial ^2}{\partial t_s^2} \lambda_{[r]}$ are non-zero eigenvalues of matrix $\matr{G}\frac{\partial ^2}{\partial t_s^2}\matr{A} \matr{G}.$
The inequality \eqref{eq: the negative-semidefinite matrix} follows from subtracting two positive terms, $-\!\sum_{i=1}^r \frac{\frac{\partial^2}{\partial t_s^2}\lambda_i}{1\!+\!2\lambda_i}\!$ and $\!2 \sum_{i=1}^r \left(\frac{\frac{\partial }{\partial t_s}\lambda_i}{1\!+\!2\lambda_i}\right)^2$. 
The term $-\!\sum_{i=1}^r \frac{\frac{\partial^2}{\partial t_s^2}\lambda_i}{1\!+\!2\lambda_i}\!$ is positive because 
\begin{multline}
    \frac{\partial^2}{\partial t_s^2}f_1=\frac{-4t(1+2(s\pow)^2t(1-2t\cdot t_s)^3)}{(1+2s\pow(1+t_s)(1-2t\cdot t_s))^3}\\
    -\frac{4s\pow t(2t+3t^2t_s^2+(1-3t\cdot t_s)^2)}{(1+2s\pow(1+t_s)(1-2t\cdot t_s))^3}
\end{multline}
and
\begin{multline}
    \frac{\partial^2}{\partial t_s^2}f_2=\frac{-4t(2+2(s\pow)^2t(1-4t\cdot t_s)^3)}{(1+2s\pow(1+t_s)(1-4t\cdot t_s))^3}\\
    -\frac{4s\pow t(4t+12t^2t_s^2+(1-6t\cdot t_s)^2)}{(1+2s\pow(1+t_s)(1-4t\cdot t_s))^3}
\end{multline}
are negative for $t>0$ and $t_s\in\left(0,\min\left\{\frac{1}{4t},1\right\}\right)$, resulting in the negative-semidefinite matrix $\matr{G}\frac{\partial ^2}{\partial t_s^2}\matr{A} \matr{G}$. Therefore, the non-zero eigenvalues of $\matr{G}\frac{\partial ^2}{\partial t_s^2}\matr{A} \matr{G}$, $\frac{\partial^2}{\partial t_s^2}\lambda_{[r]}$, are negative. 

However, the sign of the term $g^{(3)}_{2,t_s}(s,t,t_s)$ is not fixed for $t>0$ and $t_s\in \left(0,\min\left\{\frac{1}{4t},1\right\}\right)$, leading to \eqref{eq: the negative-semidefinite matrix} is not a non-decreasing function with respect to increasing $t_s$.
Consequently, to lower bound $g^{(2)}_{t_s}(s,t,t_s)$, we find a 
\begin{align}
\tilde T_s:=\arginf {t_s\in (0, \min\{1,1/4t\})} g^{(2)}_{t_s}(s,t,t_s)
\end{align}
such that  
\begin{align}
    g^{(2)}_{t_s}(s,t,\tilde T_s)\leq g^{(2)}_{t_s}(s,t,t_s),
    \label{eq: the inequality of g_b with respect to tilde T_s and T_s}
\end{align} 
for all $t_s \in\left(0,\min\left\{\frac{1}{4t},1\right\}\right)$.
% {\color{red} why did I use $\tilde T_s$ instead of $\underline T_s$? Since the }

Consequently, from \eqref{eq: initial SPA with B}, \eqref{eq: the upper bound of g}, \eqref{eq: the negative-semidefinite matrix}, and \eqref{eq: the inequality of g_b with respect to tilde T_s and T_s}, we conclude that the uniform upper bound of the PUPE for a given $t$, $s_1$, $s$, and any $\hat d_{\bar{[s]}}$ as follows:
\begin{multline}
    \mathds{E}\left[\min\left\{1,\theta(s)  \exp(t\|\tilde Z^n\|^2)\right.\right.\\
    \left.\left. \cdot \mathds{E}\left[\exp\left(-t\|B^n_{\bar {[s]}}+X^n_{[s],d_{[s]}}-\bar {X}^n_{[s],\hat d_{[s]}}+\tilde Z^n\|^2\right)\right]
    \right\}\right]\!\!\!\\
    \leq \frac{\exp\left(\bar g(s,s_1,t,\underline T_s)\right)}
    {T^*\sqrt{2\pi}}\!
    \bigg(\!g^{(2)}_{t_s}(s,t,\underline T_s)\!\bigg)^{\!-\frac{1}{2}}\!+\!O\bigg(\frac{\exp(-n)}{\sqrt{n}}\!\bigg),\label{eq: the final bound with B}
\end{multline}
where $\underline T_s \in \left(0,\min\left\{\frac{1}{4t},1\right\}\right)$ and $\bar T_s \in \left(0,\min\left\{\frac{1}{4t},1\right\}\right)$ satisfying \eqref{eq: the condition of lower bound of saddlepoint} and \eqref{eq: the condition of upper bound of saddlepoint}, respectively.
% The first inequality is because $T_s(1-T_s)>T^*,$ if $0\leq \underline T_s\leq T_s\leq \bar T_s\leq \min\left\{\frac{1}{4t},1\right\}.$
% The inequality \eqref{eq: the final error probability with B} follows from $g(s,s_1, t, T_s)<\bar g(s,s_1, t, \underline  T_s)$ for all $t_s\in \left(0,\min\left\{\frac{1}{4t},1\right\}\right)$ and $g_b(s,s_1,t,ts)\leq \frac{\partial ^2}{\partial  t_s^2} g(s,s_1 t,t_s)$. The inequality \eqref{eq: the final bound with B} follows from \eqref{eq: the inequality of g_b with respect to  tilde T_s and T_s} for all $t_s \in\left(0,\min\left\{\frac{1}{4t},1\right\}\right)$.
% To evaluate the RCU bound given in \eqref{eq: Chernoff bound; RCU bound}, we also have 

By applying the same approach to the first term in \eqref{eq: Chernoff bound; RCU bound}, we have 
\begin{multline}
    \mathds{E}\left[\min\left\{1,\theta(s)  \exp(t\|\tilde Z^n\|^2)\right.\right.\\
    \left.\left.\cdot \mathds{E}\left[\exp\left(-t\|X_{[s],d_{[s]}}-\bar {X}_{[s],\hat d_{[s]}}+\tilde Z^2\|^2\right]\right)\right\}\right]\\
    =\frac{\exp\left(g(s,t,T_s)\right)}
    {T_s(1-T_s)\sqrt{2\pi}}\left(g^{(2)}_{t_s}(s,t,T_s)\right)^{-\frac{1}{2}}+O\bigg(\frac{\exp(-n)}{\sqrt{n}}\bigg),\label{eq: the final bound without B}
\end{multline}
where $T_s\in\left(0,\min\left\{\frac{1}{4t},1\right\}\right)$ such that $g^{(1)}_{t_s}(s,t,T_s)=0$.
 
By substituting \eqref{eq: Chernoff bound; RCU bound}, \eqref{eq: the final bound with B}, and \eqref{eq: the final bound without B} into \eqref{eq: initial error event}, we complete the proof of Theorem~\ref{thm CD2}.

%% file: back/Appendix_Proof_thm2.tex
% !TeX root = ../ISIT2024.tex

\section{Proof of Theorem~\ref{thm CD with delay information}}\label{appen: proof of theorem with delay information}\vspace{-1mm}
In the following, we apply the RCU bound to express the PUPE defined in Definition~\ref{definition CD3}. 
For a given $\setd$, the PUPE is upper-bounded as follows:\vspace{-0.5mm}
\begin{multline}
    \hspace{-3mm}{\mbox{P}_{\text{PUPE}}}\leq\! p_0\! +\!\!\sum_{\sett\subseteq [\ka]}\!\frac{s\binom{\ka}{s}}{\ka} 
    \mbox{Pr}\left( \raisebox{1.6ex}{$\displaystyle\bigcup _{\substack{{\bar {X}_{[s]}\subseteq f(\mathcal M\setminus M_{[\ka]}),}\\{\hat d_{[\ka]}\in\mathfrak{S}(\setd)}}}$}\|Z^n\|^2 \right.\\
    \left.\geq\left\|\tilde Y^n-\bar {X}^n_{[s],\hat d_{[s]}}-X^n_{\bar {[s]},\hat d_{\bar {[s]}}}\right\|^2\right),\label{eq: initial error event with delay information}
\end{multline}
where $\mathfrak{S}(D^{K_a})$ is the set containing all permutations of $\setd$.

By applying the RCU bound to the tail probability in \eqref{eq: initial error event with delay information}, we have\vspace{-0.5mm}
\begin{flalign}
    &\mbox{Pr}\left( \raisebox{1.6ex}{$\displaystyle\bigcup _{\substack{{\bar {X}_{[s]}\subseteq f(\mathcal M\setminus M_{[\ka]}),}\\{\hat d_{[\ka]}\in\mathfrak{S}(\setd)}}}$}\|Z^n\|^2\geq \left\|\tilde Y^n-\bar {X}^n_{[s],\hat d_{[s]}}-X^n_{\bar {[s]},\hat d_{\bar {[s]}}}\right\|^2\right)\nonumber\\
    &\leq \mbox{E}\left[\min\left\{1,  \theta_2(s)\mbox{Pr}\bigg(\|Z^n\|^2\geq \nonumber\right.\right.\\
    &\hspace{1.5cm}\left.\left.\left. \left\| X^n_{[s], d_{[s]}}-\bar {X}^n_{[s],\hat d_{[s]}}+Z^n\right\|^2 \Big| X^n_{[\ka]},Z^n\right)\right\}\right]\nonumber\\    
    &\quad +\sum _{\substack{{\hat d_{\bar{[s]}}\in \hat{\mathfrak{S}}(\bar{[s]},\setd),}\\{s_1\neq 0}}} \mbox{E}\left[\min\left\{1,  \theta_2(s)\mbox{Pr}\bigg(\|Z^n\|^2\geq \nonumber\right.\right.\\
    &\qquad\left.\left.\left. \left\| X^n_{[s], d_{[s]}}-\bar {X}^n_{[s],\hat d_{[s]}}+B^n_{\bar{[s]}}+Z^n\right\|^2 \Big| X^n_{[\ka]},Z^n\right)\right\}\right],\!\!\!\!\!\label{eq: final with delay information}
\end{flalign}
where $\theta_2(s):=\binom{\textsfM-\ka}{s}s!$ and  $\hat{\mathfrak{S}}(\bar{[s]},\setd)$ contains all permutations of $\ka-s$ terms selected from $\setd$. The set $\hat{\mathfrak{S}}(\bar{[s]},\setd)$ only considers all permutations of $\ka-s$ terms selected from $\setd$ since the permutations of delays of $s$ incorrectly decoded codewords, i.e., $s!$, are calculated in $\theta_2(s)$.

The terms $B^n_{\bar{[s]}}$ and $s_1$ are defined in Appendix \ref{appen: proof of theorem CD2}. The inequality follows from the RCU bound. Additionally, we express the probability that the receiver matches the delays to correctly decoded codewords ($s_1=0$) and the probability that the receiver mismatches the delays to correctly decoded codewords separately.

We upper-bound \eqref{eq: final with delay information} by using the results from Theorem~\ref{thm CD2} with the term $\theta_2(s)$ modified from $\theta(s)$. Additionally, for a given $s\in[\ka]$, we upper-bound the number of permutations in the set $\hat{\mathfrak{S}}(\bar{[s]},\setd)$ satisfying a given $s_1\in[\ka-s]$ by $\binom{\ka-s}{s_1}\frac{(s+s_1)!}{s!}$, which completes the proof.

%% file: back/Appendix_Proof_lemma.tex
% !TeX root = ../ISIT2024.tex
\section{Proof of Lemma 1} \label{appendix proof of lemma}

In Lemma 1, the function $q(t_s)$ is convex by assumption, which implies $q^{(1)}(t_s)$ is a non-decreasing function with respect to  increasing $t_s \in \mathcal R$. Additionally, we have the term $T_s\in \mathcal R$ satisfies $$q^{(1)}(T_s)=\sum_{i=1}^{r}\frac{\frac{\partial }{\partial t_s}\tilde {\lambda}_i\big|_{t_s=T_s}}{1+2\tilde {\lambda}_i}.$$
By finding the functions $q_1(t_s)$ and $q_2(t_s)$ such that $$q_2(T_s) \leq \sum_{i=1}^{r}\frac{\frac{\partial }{\partial t_s}\tilde{\lambda}_i\big|_{t_s=T_s}}{1+2\tilde{\lambda}_i} \leq q_1(T_s),$$ we can find the term $\bar T_s \geq T_s $ by fulfilling $q^{(1)}(\bar T_s)=q_1(\bar T_s)$ and the term $\underline T_s \leq T_s $ by fulfilling $q^{(1)}(\underline T_s)=q_2(\underline T_s)$.

To proceed, we find the functions $q_1(t_s)$ and $q_2(t_s).$
We derive $q_1(t_s)$ and $q_2(t_s)$ for $\underline \varrho:= \min_{i\in[n]} \{\frac{\partial }{\partial t_s}A(t_s)_{i,i}\} \geq 0$, $\underline \varrho <0 <\bar \varrho:=\max_{i\in[n]} \{\frac{\partial }{\partial t_s}A(t_s)_{i,i}\}$, and $\bar \varrho<0$, separately. 

We first derive $q_1(t_s)$ for $\underline \varrho \geq 0$ as follows: 
\begin{align*}
    \sum_{i=1}^{r}\frac{\frac{\partial }{\partial t_s}\tilde{\lambda}_i}{1+2\tilde{\lambda}_i}\leq \sum_{i=1}^{r}\frac{\partial }{\partial t_s}\tilde{\lambda}_i=\mbox{Tr}\left(\matr{K} \frac{\partial \matr{A}}{\partial t_s}\right),
\end{align*}
where the first inequality is because $\tilde{\lambda}_{[r]}$ are non-zero eigenvalues of a positive-semidefinite matrix, and the second equality follows from the fact that the sum of eigenvalues is the trace of the matrix. For $\underline \varrho <0<\bar \varrho$, we have 
\begin{align*}
    \sum_{i=1}^{r}\frac{\frac{\partial }{\partial t_s}\tilde{\lambda}_i}{1+2\tilde{\lambda}_i}\leq \sum_{i\in \mathfrak{R}^+}\frac{\partial }{\partial t_s}\tilde{\lambda}_i\leq \bar \varrho \cdot \mbox{Tr}\left(\matr{K} \right),
\end{align*}
where $\mathfrak{R}^+$ is a set containing indices of positive eigenvalues. The first inequality follows $1+2\tilde{\lambda}_i\geq 1$ and the removal of negative terms. The second inequality follows from the definition of $\bar \varrho$. For $\bar \varrho\leq 0$, we have 
\begin{align*}
    \sum_{i=1}^{r}\frac{\frac{\partial }{\partial t_s}\tilde{\lambda}_i}{1+2\tilde{\lambda}_i}
    \leq \frac{\sum_{i=1}^{r} \frac{\partial }{\partial t_s}\tilde{\lambda}_i}{1+2\Lambda}
    =\frac{\mbox{Tr}(\matr{KA}(t_s))}{1+2\Lambda},
\end{align*}
where the first inequality follows the definition of $\Lambda$. The second equality follows because the sum of eigenvalues is the matrix trace. Therefore, we have
    \begin{align*}
        q_1(t_s):=&
        \begin{cases}
          \mbox{Tr}(\matr{K} \frac{\partial \matr{A}}{\partial t_s}), &\text{if } \underline\varrho  \geq 0\\
          \bar \varrho\cdot \mbox{Tr}(\matr{K }),  & \text{if } \underline \varrho <0 <\bar \varrho\\
           \mbox{Tr}(\matr{K} \frac{\partial \matr{A}}{\partial t_s})/(1+2\Lambda),  & \text{o.w., }
        \end{cases}
    \end{align*}
By using the same approach, we have 
    \begin{align*}
        q_2(t_s):=&
        \begin{cases}
          \mbox{Tr}(\matr{K} \frac{\partial \matr{A}}{\partial t_s})/(1+2\Lambda), &\text{if } \underline\varrho  \geq 0\\
          \underline \varrho\cdot \mbox{Tr}(\matr{K }),  & \text{if } \underline \varrho <0 <\bar \varrho\\
           \mbox{Tr}(\matr{K} \frac{\partial \matr{A}}{\partial t_s}),  & \text{o.w., }
        \end{cases}
    \end{align*}
Consequently, we conclude the conditions of $\bar T_s$ and $\underline T_s$ are $q^{(1)}(\bar T_s)=q_1(\bar T_s)$ and 
    $q^{(1)}(\underline T_s)=q_2(\underline T_s)$, respectively, which completes the proof.

%% file: AUMAC_arXiv.bbl
\begin{thebibliography}{10}
\providecommand{\url}[1]{#1}
\csname url@samestyle\endcsname
\providecommand{\newblock}{\relax}
\providecommand{\bibinfo}[2]{#2}
\providecommand{\BIBentrySTDinterwordspacing}{\spaceskip=0pt\relax}
\providecommand{\BIBentryALTinterwordstretchfactor}{4}
\providecommand{\BIBentryALTinterwordspacing}{\spaceskip=\fontdimen2\font plus
\BIBentryALTinterwordstretchfactor\fontdimen3\font minus \fontdimen4\font\relax}
\providecommand{\BIBforeignlanguage}[2]{{%
\expandafter\ifx\csname l@#1\endcsname\relax
\typeout{** WARNING: IEEEtran.bst: No hyphenation pattern has been}%
\typeout{** loaded for the language `#1'. Using the pattern for}%
\typeout{** the default language instead.}%
\else
\language=\csname l@#1\endcsname
\fi
#2}}
\providecommand{\BIBdecl}{\relax}
\BIBdecl

\bibitem{lin2023information}
S.-C. Lin, T.-H. Chang, E.~Jorswieck, and P.-H. Lin, \emph{Information theory, mathematical optimization, and their crossroads in 6G system design}.\hskip 1em plus 0.5em minus 0.4em\relax Springer, 2023.

\bibitem{ElGamal.2011}
A.~El~Gamal and Y.-H. Kim, \emph{Network Information Theory}.\hskip 1em plus 0.5em minus 0.4em\relax Cambridge University Press, 2011.

\bibitem{Polyanskiy_perspective_on_UMAC}
Y.~Polyanskiy, ``A perspective on massive random-access,'' in \emph{2017 IEEE International Symposium on Information Theory (ISIT)}, 2017, pp. 2523--2527.

\bibitem{Guo_many_access_asymptotic_capacity}
X.~Chen, T.-Y. Chen, and D.~Guo, ``Capacity of {Gaussian} many-access channels,'' \emph{IEEE Transactions on Information Theory}, vol.~63, no.~6, pp. 3516--3539, 2017.

\bibitem{Effros_RAC}
R.~C. Yavas, V.~Kostina, and M.~Effros, ``Random access channel coding in the finite blocklength regime,'' \emph{IEEE Transactions on Information Theory}, vol.~67, no.~4, pp. 2115--2140, 2021.

\bibitem{Effros_RAC_MAC}
------, ``{Gaussian} multiple and random access channels: Finite-blocklength analysis,'' \emph{IEEE Transactions on Information Theory}, vol.~67, no.~11, pp. 6983--7009, 2021.

\bibitem{T_fold_1}
O.~Ordentlich and Y.~Polyanskiy, ``Low complexity schemes for the random access {Gaussian} channel,'' in \emph{2017 IEEE International Symposium on Information Theory (ISIT)}, 2017, pp. 2528--2532.

\bibitem{AMUC1981}
T.~Cover, R.~McEliece, and E.~Posner, ``Asynchronous multiple-access channel capacity,'' \emph{IEEE Transactions on Information Theory}, vol.~27, no.~4, pp. 409--413, 1981.

\bibitem{Cyclic_decoding}
H.~Ebrahimzadeh~Saffar, M.~Badiei~Khuzani, and P.~Mitran, ``Time-asynchronous {G}aussian multiple access relay channel with correlated sources,'' \emph{IEEE Transactions on Information Theory}, vol.~62, no.~1, pp. 309--321, 2016.

\bibitem{Polyanskiy_low_complexity_AUMAC}
K.~Andreev, S.~S. Kowshik, A.~Frolov, and Y.~Polyanskiy, ``Low complexity energy efficient random access scheme for the asynchronous fading {MAC},'' in \emph{2019 IEEE 90th Vehicular Technology Conference (VTC2019-Fall)}, 2019, pp. 1--5.

\bibitem{Polyanskiy_Short_packet}
S.~S. Kowshik, K.~Andreev, A.~Frolov, and Y.~Polyanskiy, ``Short-packet low-power coded access for massive {MAC},'' in \emph{2019 53rd Asilomar Conference on Signals, Systems, and Computers}, 2019, pp. 827--832.

\bibitem{Guo_CS}
X.~Chen, L.~Liu, D.~Guo, and G.~W. Wornell, ``Asynchronous massive access and neighbor discovery using {OFDMA},'' \emph{IEEE Transactions on Information Theory}, vol.~69, no.~4, pp. 2364--2384, 2023.

\bibitem{Group_testing_delays}
S.~Derya, S.-C. Yu, H.-W. Young, E.~A. Jorswieck, P.-H. Lin, and S.-C. Lin, ``Joint delay and user activity detection in asynchronous massive access,'' in \emph{2024 33rd Wireless and Optical Communications Conference (WOCC)}, 2024, pp. 175--179.

\bibitem{ISIT_worst_case}
J.-S. Wu, P.-H. Lin, M.~A. Mross, and E.~A. Jorswieck, ``Worst-case per-user error bound for asynchronous unsourced multiple access,'' in \emph{{IEEE} International Symposium on Information Theory, {ISIT} 2024, Athens, Greece, July 7-12, 2024}.\hskip 1em plus 0.5em minus 0.4em\relax {IEEE}, 2024, pp. 3207--3212.

\bibitem{Polyanskiy.2010}
Y.~Polyanskiy, H.~V. Poor, and S.~Verdu, ``Channel coding rate in the finite blocklength regime,'' \emph{IEEE Trans. Inf. Theory}, vol.~56, no.~5, pp. 2307--2359, 2010.

\bibitem{Saddlepoint_approximation_keystone}
J.~Jensen, \emph{\BIBforeignlanguage{English}{Saddlepoint approximations}}.\hskip 1em plus 0.5em minus 0.4em\relax Clarendon Press Oxford, 1995, oxford Science Publications.

\bibitem{saddlepoint_of_RCUs}
A.~Martinez and A.~G. i~Fàbregas, ``Saddlepoint approximation of random-coding bounds,'' in \emph{2011 Information Theory and Applications Workshop}, 2011, pp. 1--6.

\bibitem{Quadratic_form}
S.~Provost and A.~Mathai, \emph{Quadratic Forms in Random Variables: Theory and Applications}, ser. Statistics : textbooks and monographs.\hskip 1em plus 0.5em minus 0.4em\relax Marcel Dekker, 1992.

\bibitem{Horn_Johnson_1985}
R.~A. Horn and C.~R. Johnson, \emph{Matrix Analysis}.\hskip 1em plus 0.5em minus 0.4em\relax Cambridge University Press, 1985.

\end{thebibliography}
